# Importance of Explicit Vectorization for CPU and GPU Software Performance

Neil G. Dickson     Kamran Karimi     Firas Hamze

D-Wave Systems Inc.
100-4401 Still Creek Drive
Burnaby, British Columbia
Canada, V5C 6G9
{ndickson, kkarimi, fhamze}@dwavesys.com

Much of the current focus in high-performance computing is on multi-threading, multi-computing, and graphics processing unit (GPU) computing. However, vectorization and non-parallel optimization techniques, which can often be employed additionally, are less frequently discussed. In this paper, we present an analysis of several optimizations done on both central processing unit (CPU) and GPU implementations of a particular computationally intensive Metropolis Monte Carlo algorithm. Explicit vectorization on the CPU and the equivalent, explicit memory coalescing, on the GPU are found to be critical to achieving good performance of this algorithm in both environments. The fully-optimized CPU version achieves a 9x to 12x speedup over the original CPU version, in addition to speedup from multi-threading. This is 2x faster than the fully-optimized GPU version.

## 1  Introduction

It is common to examine performance increases from parallelism in the form of CPU and GPU multi-threading [1] [2], but vectorization and non-parallel optimization techniques remain less common in the literature. Based on theoretical peak performance values or performance test results, it is sometimes concluded or even assumed that GPU computation is significantly faster than CPU computation [3] [4]. However, explicit optimization of the CPU code, a process by which a programmer manually intervenes to maximize performance, is often not considered. Due to several well-documented GPU performance factors [5], it is common for GPU developers to become involved in many aspects of coding that can positively affect performance. CPU optimization, however, is usually left to the compiler, and it has been argued that compilers achieve high utilization of the CPU's computing power [6] [7].

The goal of this paper is to present a performance analysis of CPU and GPU implementations of a specific Metropolis Monte Carlo algorithm at different levels of optimization, as outlined in Table 1.

| Impl. | CPU/GPU | Multi-Threaded | Compiler Optimization Enabled | Basic Optimizations (Section 2) | Vectorized MT19937 & Flipping (Section 3) | Vectorized Data Updating (Sections 3.1 & 3.2) |
|---|---|---|---|---|---|---|
| A.1a | CPU | ✓ | | | | |
| A.1b | CPU | ✓ | ✓ | | | |
| A.2a | CPU | ✓ | | ✓ | | |
| A.2b | CPU | ✓ | ✓ | ✓ | | |
| A.3  | CPU | ✓ | ✓ | ✓ | ✓ | |
| A.4  | CPU | ✓ | ✓ | ✓ | ✓ | ✓ |
| B.1  | GPU | ✓ | ✓ | ✓ | | |
| B.2  | GPU | ✓ | ✓ | ✓ | ✓ | ✓ |

Table 1. Implementations at different levels of optimization

Vector instruction sets, present on modern commodity CPUs since 2001 (referred to here as SSE), allow [8] the same operation to be executed on multiple (2, 4, 8, or 16) adjacent data elements at once, but are not frequently used explicitly. High-quality vectorization of our algorithm requires knowledge about the semantics of the algorithm, which a compiler cannot (and arguably, should not) assume. Software developers, on the other hand, must already have knowledge of these semantics in order to design and implement the algorithm.

For example, the MT19937 (pseudo)random number generation algorithm [9] keeps an array of 624 numbers. An alternative implementation, used in this paper, keeps 4×624=2,496 numbers and uses SSE to generate 4 random numbers in roughly the same time as each random number before. The resulting random number sequence would be different, so in order for a compiler to safely replace the algorithm with something not equivalent to the source code, it must first determine that the sole purpose of the code is to generate many random numbers. An analogous scenario would be a compiler recognizing code performing bubble sort, a stable sort, determining that the sorting stability is not necessary, and then replacing the code with heap sort, an unstable sort [10].

The application we consider involves an Ising model [11], which is a system of spin variables $s_i$, i.e. variables that can only be +1 or -1, with a cost function of the form:

$$f(s_0, \ldots, s_{n-1}) = -\sum_i h_i s_i - \sum_{i,j} J_{ij} s_i s_j$$

where $h_i$ and $J_{ij}$ are the values that describe the particular Ising model. The objective of a Metropolis algorithm [12] operating on such a system is usually to sample from a Boltzmann distribution [13] over the possible values of spins, for example, to minimize the cost [14]. The general Metropolis sweep algorithm used in this paper can be summarized as in Figure 1.

```
for each spin i,
    if uniform(0,1) random number < probability of flipping the sign of i,
        flip the sign of i
        for each spin j adjacent to i,
            update data used to find probability of flipping j
```

Figure 1. A Metropolis Monte Carlo sweep of an Ising model

The optimized implementations were developed in a Quantum Monte Carlo (QMC) [15] simulation context and use Parallel Tempering, as explained in [16] and [17]. For the purposes of this paper, though, the most relevant information is that each spin is adjacent to 6, 7, or 8 other spins, and that millions of the Metropolis sweeps shown in Figure 1 must be performed on millions of systems, each with up to 32,768 spins. With these high computational requirements, even the best-optimized implementations require months of computation time on thousands of multi-core computers. For densely-connected Ising models, a different vectorized implementation than the ones described in Section 3 would be advised, since the sparseness is exploited by our implementations.

Explicit CPU vectorization has shown impressive results in other contexts [18] [19]. However, those performance comparisons are relative to implementations where the compiler does not implicitly vectorize the code, so improvement over a compiler's implicit vectorization may not be known for those cases. It has also been found in other contexts that a GPU implementation is not guaranteed to surpass a CPU implementation in performance [20]. Again, many factors could explain such a discrepancy, so we attempted to address these concerns by exploring a number of potential optimizations applicable to our context.

The rest of the paper is organized as follows. Section 2 explains a number of non-parallel optimization techniques that can be applied to both the CPU and the GPU implementations. Section 3 shows how different parts of the code were vectorized. This section also explains how memory coalescing for GPU was performed. Section 4 presents the results of a number of performance tests we performed using CPU and GPU code at different levels of optimization. Section 5 concludes the paper.

# 2 Basic Optimizations

The optimizations presented here focus only on the Metropolis sweep algorithm of Figure 1, as everything else remains largely unchanged, including multi-threading, which is explained in [16]. The basic optimization techniques we used on this algorithm include: branch elimination, simplification of data structures, result caching, and approximation.

Apart from the approximation, these optimizations simplified the code (e.g. in readability and size) and took less than a day to implement and test. The approximation, 3 lines of non-obvious code, took roughly three days to design, implement, test for accuracy, and document. Our GPU implementation uses all of these optimizations. Many other GPU optimizations were tried in the several weeks spent trying to improve the GPU performance, but only one resulted in a non-negligible improvement, and is described in Section 3.2.

## 2.1 Branch Elimination

Commodity CPUs have had the ability to run parts of multiple instructions at the same time (known as superscalar architecture) since 1993 [8]. The drawback of this feature is that if the current instruction is a conditional branch, for example due to an if-statement, either the CPU cannot start running any instructions past that branch, or it must make a guess as to what the next instruction will be. This guessing is known as branch prediction. When the guess is wrong (a misprediction), several partially-completed instructions may need to be removed from the CPU's pipeline, which can cause stalls [21]. Eliminating unnecessary branches can alleviate this problem.

This optimization had a large impact on the performance, and combined with the data structure simplification, made the code simpler. On its own, the resulting code was smaller, but admittedly less readable. In the original code, shown in Figure 2, the inner loop of the Metropolis sweep contained two frequently-mispredicted branches. Note that the names of the data structures and their specific purposes are not relevant to the discussion.

```
for (edge_index=0; edge_index<num_local_edges; edge_index++) {
    curr_edge = incident_edges[edge_index];

    if (graph_edges[curr_edge][0]==curr_spin)
        curr_nbr = graph_edges[curr_edge][1];
    else
        curr_nbr = graph_edges[curr_edge][0];

    if (p_in_model->isATauEdge[curr_edge])
        h_eff_tau[curr_nbr] -= 2*S_mul*J[curr_edge];
    else
        h_eff_space[curr_nbr] -= 2*S_mul*J[curr_edge];
}
```
Figure 2. Original loop updating data for spins adjacent to one that has just flipped

However, both branches can be eliminated, as in Figure 3.

```
for (edge_index=0; edge_index<num_local_edges; edge_index++) {
    curr_edge = incident_edges[edge_index];

    neighbours = graph_edges[curr_edge];
    curr_nbr = neighbours[neighbours[0]==curr_spin];

    h_effective = p_in_model->isATauEdge[curr_edge]? h_eff_tau : h_eff_space;
    h_effective[curr_nbr] -= 2*S_mul*J[curr_edge];
}
```
Figure 3. Loop updating data for spins after a flip, without using branches

Note that the value of `(neighbours[0]==curr_spin)` is 1 when true and 0 when false, so the new code still selects the neighbour of the edge that is not the current spin, to update it after flipping the current spin. Also, for x86-compatible processors since the Pentium Pro in 1995, the ternary operator (i.e. condition ? valueIfTrue : valueIfFalse) is implemented with a conditional move instruction instead of a branch when both possible values do not involve computation [21].

To further speed up this computation and avoid both of these confusing ways of accessing the data, the data structures require a more suitable design, examined in the following section.

## 2.2 Simplification of Data Structures

The data structures mentioned above are used elsewhere in the code and suffered from similar issues, as they have similar access patterns. The original graph data structure had a complex layout in memory, as shown in Figure 4.

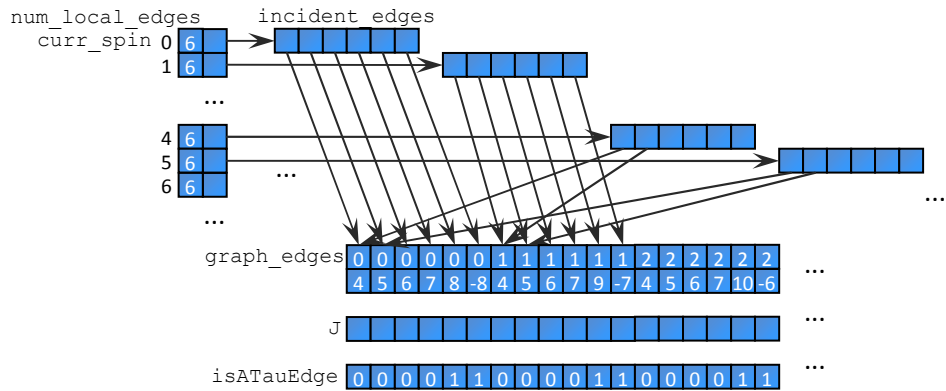

Figure 4. Original memory layout of the graph data structure representing connections between spins

However, just by "eliminating the middle man", placing J values with the spin indices they apply to, and removing `isATauEdge` (as described in the next paragraph), this simplifies to Figure 5.

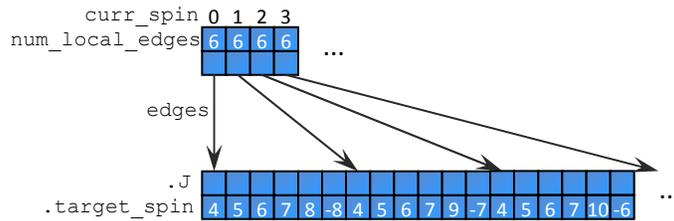

Figure 5. Simplified memory layout of the graph data structure

It happens that by design, there are always exactly two edges of each spin for which `isATauEdge` is true in these QMC simulations. Since edges can appear in any order, reordering them ahead of time such that these edges always appear after the edges for which `isATauEdge` is false allows the elimination of `isATauEdge`. The entire loop can be reduced to Figure 6.

```
for (edge=0; edge<num_local_edges-2; edge++) {
    h_eff_space[edges[edge].target_spin] -= 2*S_mul*edges[edge].J;
}
h_eff_tau[edges[edge  ].target_spin] -= 2*S_mul*edges[edge  ].J;
h_eff_tau[edges[edge+1].target_spin] -= 2*S_mul*edges[edge+1].J;
```
Figure 6. Loop updating data for spins after a flip, using a simpler data structure

The code is clearer than before and the inside of the loop is now just one line (down from 9). In addition to simply computing less, there is now less memory use, more sequential access, and fewer arrays being read in quick succession, improving memory cache use [21]. As such, this had a large performance impact on top of the branch elimination.

## 2.3 Result Caching

This is a common optimization technique. As in dynamic programming, simply avoid computing the same value multiple times, by computing it once and saving the result. Here, it doesn't appear prominently, but caching certain data in the algorithm did improve performance. For example, `(2*S_mul)` appears multiple times in Figure 6, including inside the for-loop. `S_mul` does not change within the loop and is never read without being doubled, so it can simply be doubled once, before the loop.

This improved performance slightly, but noticeably, which was surprising to us, as this is an optimization that a compiler can easily recognize and perform.  Choosing not to cache computations that are clearly duplicated may be advantageous in certain situations, such as when registers are fully occupied and the computation is sufficiently brief, but this was evidently not such a case.

As a side note, a way of indirectly caching this multiplication over larger timescales would be to instead multiply all of the $J$'s by 2 ahead of time.  More subtle forms of result caching are also possible.  For example, we generate many random numbers at a time, avoiding the overhead of cache misses and extra condition checks from frequently switching between generating random numbers and flipping spins.

### 2.4 Approximating an Exponential

In Metropolis Monte Carlo, computing the probability of flipping a spin's sign requires an exponential to be computed, which is an expensive operation, requiring roughly 83 clock cycles on our test CPU (see Section 4 for CPU information).  We created a rough approximation to compute $e^x$ in 4 clock cycles, and a more accurate approximation takes 11 clock cycles.  The more accurate approximation, with maximum relative error of roughly 1% and average relative error near zero, is generated as in Figure 7.  More detail is provided in Appendix: Derivation of Exponential Approximation.

```
1. Start with x as a 32-bit floating-point value, which must be in the range
   (-31.5 ln 2) ≤ x < (32 ln 2)
2. Multiply x by 2^25 log_2 e
3. Convert x to a 32-bit integer value
4. Add (127)2^23 (i.e. 0x3F800000) to x
5. Pretending that x is now a 32-bit floating-point value, multiply x by 2ln^2 2
6. Compute the approximate 4th root of x
```
Figure 7. An algorithm for approximating $e^x$

Note that the 11 clock cycles includes special masking to produce 0.0 for all $x < $ (-31.5 ln 2) and at least 1.0 for $x$>0, and the approximate 4th root can be computed using the approximate reciprocal-square-root instructions provided on modern Intel and AMD CPUs [8].  It was important that this approximation does not use lookup tables, so that it can also be vectorized, i.e. to compute 4 approximate exponentials at once.

The faster, less accurate approximation skips the bounds checking (the valid range is (-126 ln 2) ≤ $x$ < (128 ln 2)), reduces the factor in step 2 to $2^{23}\log_2 e$, and skips step 6.  This faster approximation was used in the performance tests for all implementations with these basic optimizations.  It is equivalent to a linear interpolation between exact values at the points where $e^x$ is a power of two, scaled by $2\ln^2 2$.  If the 4th root was exact, the more accurate approximation would be equivalent to a linear interpolation between exact values at the points where $e^{4x}$ is a power of two, scaled by $2\ln^2 2$.

## 3 Vectorization

After implementing the optimizations of Section 2, we observed that a majority of CPU time was being spent generating the large volume of random numbers needed, using the Mersenne Twister 19937 [9] algorithm to ensure high quality of randomness.  Nearly a 4x speedup of the random number generation can be achieved by always generating 4 random numbers together using SSE. This is accomplished by interlacing 4 MT19937 random number generators with different seeds, working in parallel. A pair of lines from MT19937 comes in Figure 8.

```
y = (data[i]&UPPER_MASK)|(data[i+1]&LOWER_MASK);
data[i] = data[i+397] ^ (y>>1) ^ ((y&1) ? MASK_A : 0);
```
Figure 8. Two example lines of MT19937 random number generation algorithm

The ANDs, XORs, OR, logical shift right, and ternary operator in Figure 8 can all be performed on 4 adjacent 32-bit integers concurrently, so one can conceptually just change the type of `data` and `y` from single 32-bit integers to quadruplets of adjacent 32-bit integers. Then, the operations to generate 4 random numbers would be almost identical to those needed to generate 1. Figure 9 depicts the first line of Figure 8 in scalar and vector forms. Figure 10 depicts the vector version of the ternary operator, described more thoroughly in [18], since its particular operations differ from its scalar version.

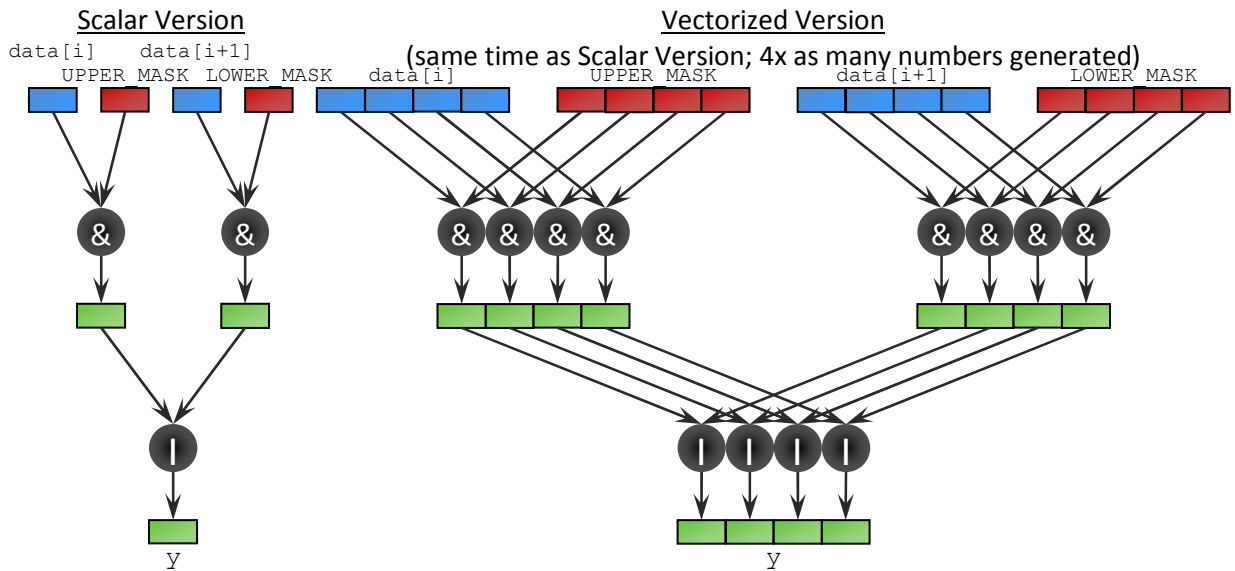

Figure 9. Graphical depiction of operations in scalar and vector versions of a line of Figure 8

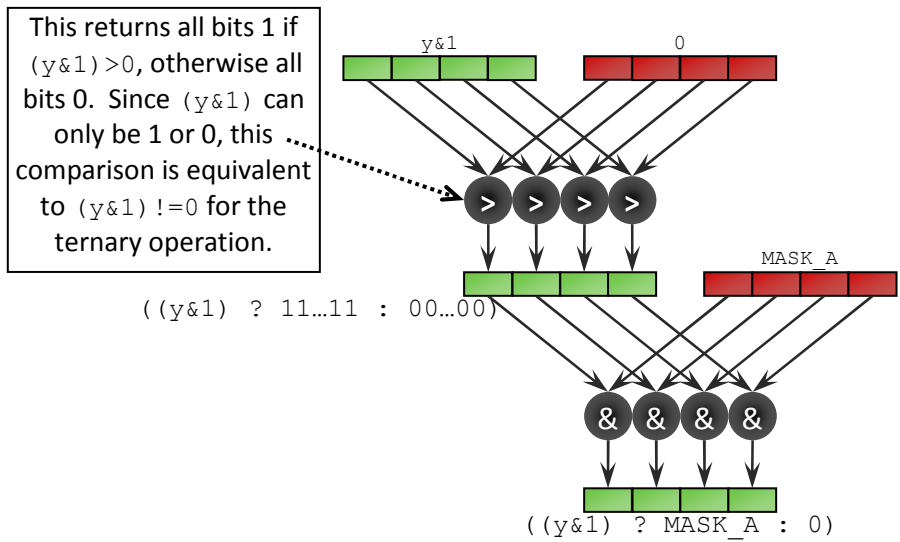

Figure 10. Graphical depiction of the vectorized version of the ternary operator in Figure 8

With random vectors generated, the question is whether they can be used in vector form as well. If 4 independent spin systems are interlaced, or if spins are reordered such that each quadruplet of spins has no edge between them, the probabilities of flipping each of 4 spins can be computed in vector form and the flipping can be done in vector form (masked by whether each spin actually flipped, similar to Figure 10). This reordering is described in Section 3.1 below. A larger-scale reordering is also used for memory coalescing on the GPU in Section 3.2.

C++ compilers do not yet natively provide operators on 128-bit data types, instead only providing compiler intrinsics to effectively write the equivalent assembly language code as if the instructions were C++ functions. As such, the vectorized versions of our Metropolis sweep algorithm are implemented directly in assembly language, which took a week to implement and test. This is a stumbling block to widespread use of vectorization, especially because different compilers have different names for the intrinsics. Having more support in C or C++ for vector operations could enable more programmers to use vectorization. An extension to simplify vectorization in C++ is presented in [18].

Note that in order to give the compiler a better opportunity to implicitly vectorize the random number generation, while still observing its regular behaviour for implementations A.1a and A.1b, implementations A.2a and A.2b use 4 random number generators interlaced, as described in this section. In that code, the operations are each performed 4 separate times in close succession, once on each generator, to allow this behaviour to be indentified more easily by a compiler. This provided some speedup, as reported in Section 4.

## 3.1 Full Vectorization on the CPU

If all 4 spins within a quadruplet are topologically identical (as described below, e.g. 4 identical independent systems), the loop that updates data after spin flips can also be vectorized. However, simulating 4 independent systems in this manner would make other components of our simulation much less efficient (e.g. the Parallel Tempering must be able to swap out the states of these systems independently of each other), so this is not done in our QMC simulations.

That said, the structure of our QMC simulations can still be exploited in order to vectorize the inner loop of the Metropolis sweep in Figure 1. All of our simulated Ising models consist of many (≥64) identical copies of a smaller Ising model, with edges connecting corresponding spins in adjacent layers, with a wrap-around from the last to the first layer, as shown in Figure 11 and Figure 12.

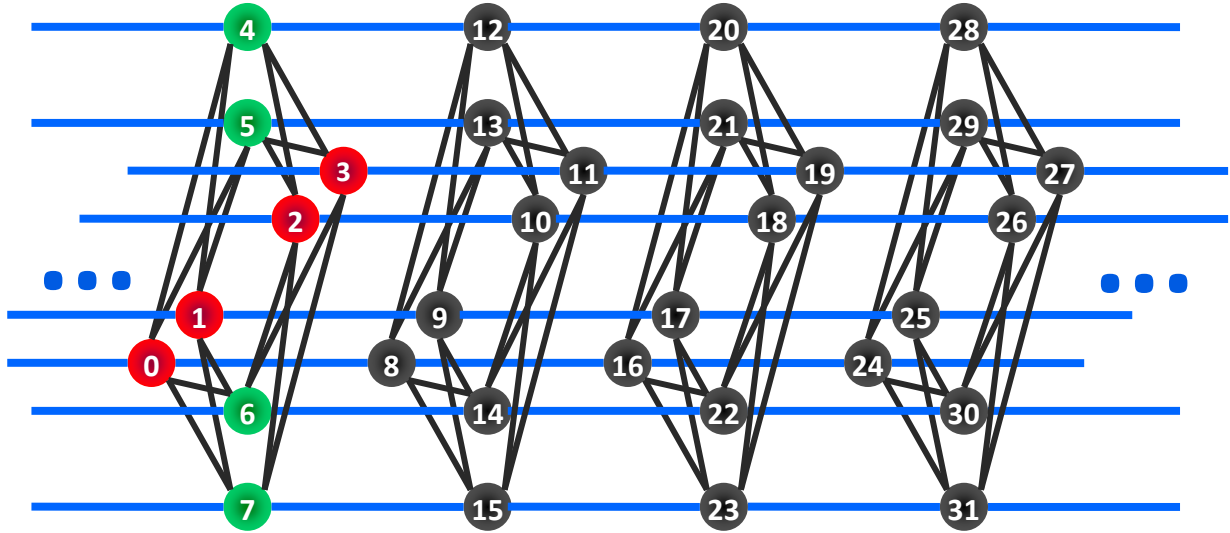

Figure 11. Section of a small example QMC Ising model, showing 4 layers each with 8 spins (quadruplet 0 of original spin order marked in red; quadruplet 1 of original spin order marked in green)

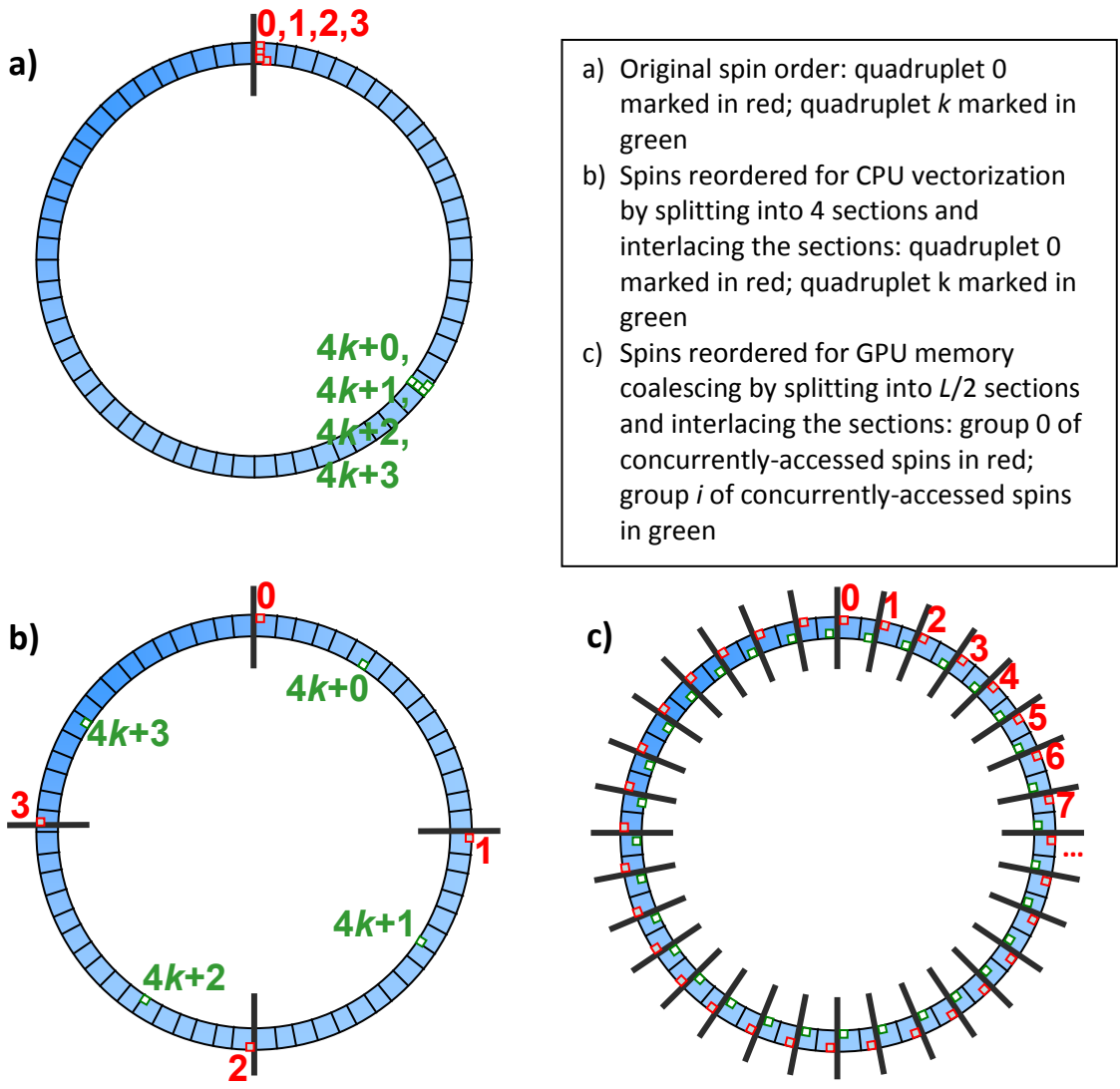

Figure 12. Reordering of spins for small example QMC Ising model with $L=64$ identical layers

Although the first quadruplet marked in Figure 11 happens to contain only spins that are not adjacent to each other (and this is not true for all of our simulations), if spin 0 and spin 1 were both flipped together, they could both attempt to update data for spin 4 at the same time, producing unpredictable results. Reordering the spins by splitting the layers into 4 sections and interlacing those sections produces the reordering in Figure 12b. With this ordering, not only is there no threat of two spins in the same quadruplet trying to update the same spin at the same time, but they also always update spins that form another quadruplet, except when an update wraps around between the first and last layers. For example, flipping spins (0,1,2,3) would require updating data for spins (16,17,18,19) identically (the spins corresponding with spin 4 before), since the layers are identical.

Because this update involves identical operations on adjacent data (that can be masked out for spins that do not flip, similar to in Figure 10), these operations can be done together in vector form. The first and last layers can be treated as a special case to handle wrapping. Note that if the number of layers is not a multiple of 4, either extra spins must be added in such a way that they can be ignored, or the remainder of layers must be left non-vectorized.

## 3.2  Memory Coalescing on GPU

Vector instruction sets like SSE are not available on current GPUs. However, current NVIDIA GPUs access memory much faster than they otherwise would when groups of 32 GPU threads, called "warps", access adjacent data concurrently, an effect known as "memory coalescing" [3]. As such, reordering the spins in a manner similar to that used to fully vectorize the CPU implementation (except using 32-way interlacing instead of 4-way interlacing), would lead to this improved memory access pattern. This is similar to having a vector processor with 32 elements in each vector, except that on the CPU, data accessed during vector operations must be in adjacent memory locations, whereas the GPU supports accessing one or more of the elements from elsewhere in memory, though with degraded performance.

To maximize the number of concurrent threads accessing adjacent data, we split the model into groups of 2 layers and interlaced all of these groups, as in Figure 12c. In that example, if not properly coordinated, a thread flipping spin 0 and a thread flipping spin 1 may attempt to update data for spin 256, the corresponding spin in the layer between, at the same time. As such, we have each thread attempt flipping spins in its even layer while only updating data to its left, then after all threads have completed their flipping, each thread updates data to its right. Then, flips are attempted for spins in the odd layers in the same manner. For a system with 256 layers, we would perform 128-way interlacing, allowing 128 GPU threads to operate on a single Ising model at once, with as many as possible operating on adjacent data.

Our GPU implementation having only basic optimizations (B.1) also has 128 GPU threads operating on each Ising model with 256 layers in the exact same manner, simply without reordering the spins first. The GPU version of the code has a random number generator for each GPU thread [16]. As with the CPU vectorized version, the GPU version with memory coalescing interlaces the 128 random number generators. Interlacing the random number generators was implemented simply by swapping the order of two array indices.

# 4  Results and Comparison

As described above, the relative performance of the CPU and GPU implementations at different levels of optimization was examined.

The system used for the CPU performance testing:
- Intel Core i7-965 Extreme (4 physical cores 2x Hyper-Threaded, so 8 logical cores at 3.2 GHz)
- 12 GB of 1066MHz RAM
- Windows Vista Ultimate 64-bit
- Visual C++ 2008 64-bit compiler, Macro Assembler (x64) Version 8.00

The system used for the GPU performance testing:
- NVIDIA GTX-285 (240 cores at 648 MHz with 1 GB of 1242 MHz RAM)
- Running CUDA driver version 196.21
- Windows 7 Ultimate 64-bit

For the GPU tests, we observed a low CPU utilization (below 1%), making the results largely independent of the test computer's CPU performance.

For the performance results below, 30,000 Metropolis sweeps were performed on 115 Ising models, each with 24,576 spins (256 layers of 96 spins), for a total of 2,826,240 spins. Data was transferred to the GPU only once, and as a result, the observed data transfer time was negligible for the GPU runs. CPU runs were performed on 1, 2, 4, 6, and 8 cores solely for comparison between CPU and GPU. As previously mentioned, multi-threading of this application is covered in [16]. All runs were performed 10 times and averaged to ensure consistency. The error bars are too small to see due to low variance in the times. The reference point for Figure 13, the average time for our original CPU code to execute on 1 core, was 5705.27 seconds (1 hour 35 minutes 5.27 seconds).

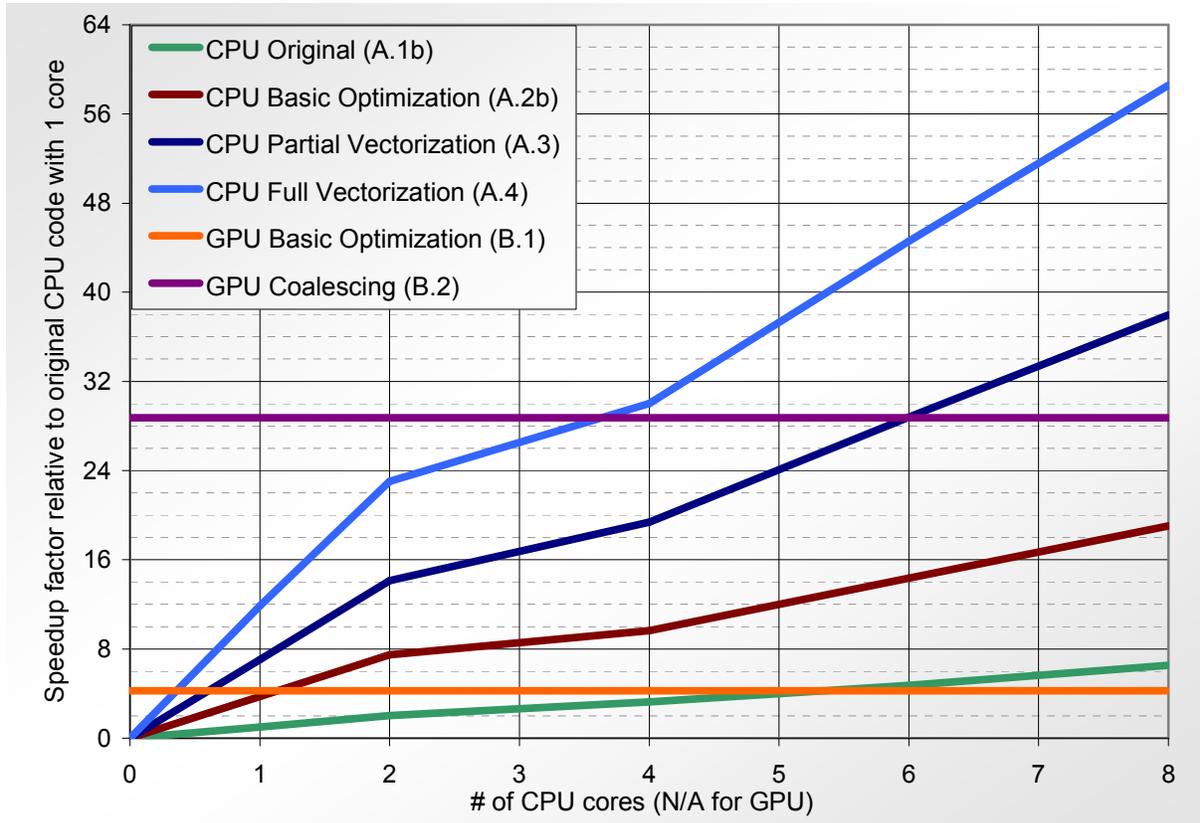

Figure 13. Relative performance at different levels of optimization of CPU code and CUDA code on GPU

The basic optimizations improved performance on the CPU by 2.91x to 3.75x, depending on the number of cores. Full vectorization contributed another factor of 3.08 to 3.16, giving a total speedup of 8.95x to 11.86x from manual optimization of the CPU code. The speedup is significant enough that using 1 core of the fully vectorized version outperforms 8 cores of the original version by 1.8x.

Reordering the spins and the random number generators to allow for memory coalescing on the GPU gave a dramatic 6.78x speedup over the GPU code with just basic optimizations, stressing the importance of using the GPU in a manner similar to a vector processor. This reorganization of memory was the only difference between the two GPU versions, so the code of both B.1 and B.2 are almost identical. Our GPU implementation with memory coalescing performed roughly as well as 4 CPU cores of our fully vectorized CPU version. Running these best-optimized implementations, the full 8 cores of the Core i7 outperform the GTX-285 by a factor of 2.04.

One source of slowdown for the GPU is the main if-statement of the Monte Carlo sweep algorithm (Figure 1). If even one thread in a warp of 32 threads flips a spin, all threads in that warp must wait for that spin to be flipped [16]. The probability of waiting for a flip in each of the 115 Ising models simulated is shown in Figure 14.

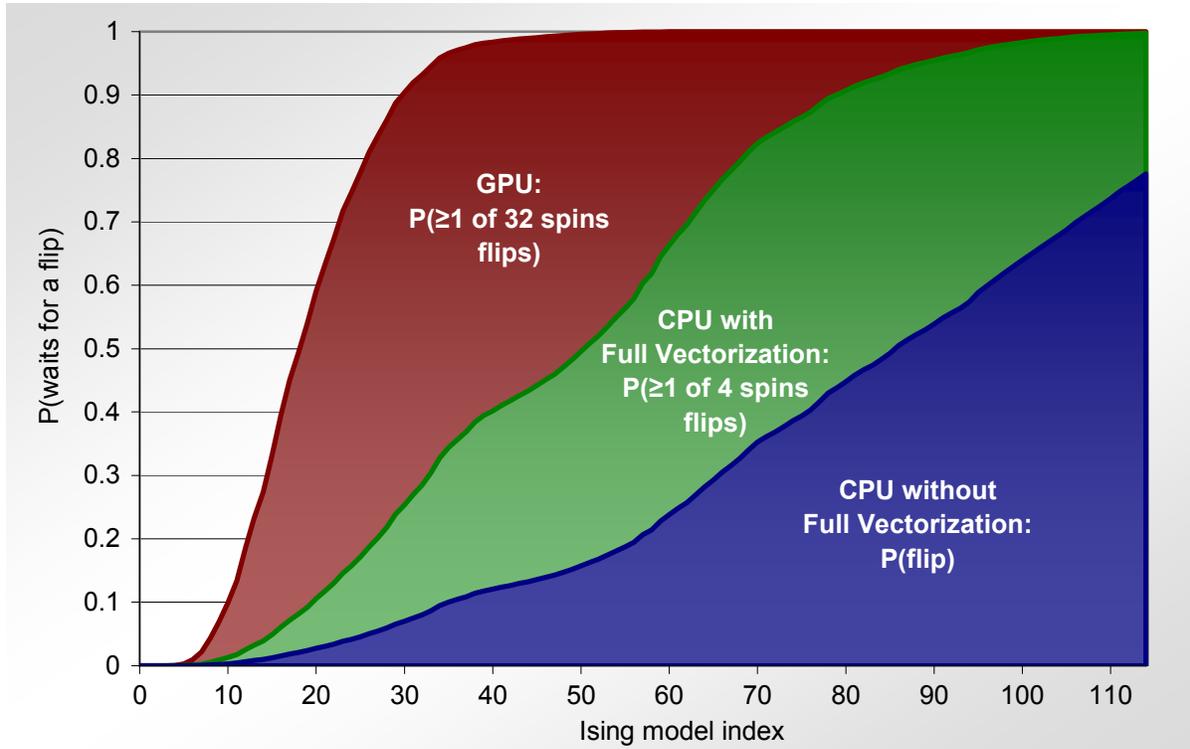

Figure 14. Probabilities of having to wait for a spin flip in different application versions

The Ising models with lower indices have a significantly lower probability of flipping a spin than those with higher indices, since they represent lower effective temperatures. The percentage of time that the A.1 CPU application must wait for the main if-statement to finish is exactly the average percentage chance of flipping, or 28.6%. However, the GPU must wait on average 82.8% of the time, i.e. whenever at least 1 spin out of 32 spins flips, which is 2.9x more often than the CPU. The A.4 CPU application must also wait more often at 56.8%, i.e. whenever at least 1 spin out of 4 spins flips, or 2.0x more often than before. On the other hand, the reason that this increase in probability of waiting for the main if-statement occurs in the first place is that the if-statement can run for multiple spins concurrently. Therefore, this does not explain the relatively low performance of the GPU, only why memory coalescing does not provide a larger speedup than the 7x observed.

Comparing the GTX-285 GPU and the Core i7 CPU based on the number of cores is not a meaningful comparison in this context, as the GPU's 240 cores are outperformed by the CPU's 8 logical cores when both are running optimized code. This discrepancy in per-core performance can be partly explained by the 4.94x difference in core frequency and the fact that each CPU core operates on 4-element vectors, whereas each GPU core operates on scalars. The remaining discrepancy could be explained by a difference in average number of clock cycles per instruction and the lower probability of waiting for a flip on the CPU.

To examine the CPU implementations more thoroughly, we ran the performance test, described above, on two implementations with compiler optimization disabled (A.1a and A.2a).

|      | A.1a  | A.1b  | A.2a  | A.2b  | A.3    | A.4    |
|------|-------|-------|-------|-------|--------|--------|
| A.1a | 1.000 | 1.508 | 1.921 | 5.652 | 10.636 | 17.886 |
| A.1b | 0.663 | 1.000 | 1.274 | 3.748 | 7.053  | 11.860 |
| A.2a | 0.521 | 0.785 | 1.000 | 2.942 | 5.537  | 9.311  |
| A.2b | 0.177 | 0.267 | 0.340 | 1.000 | 1.882  | 3.165  |
| A.3  | 0.094 | 0.142 | 0.181 | 0.531 | 1.000  | 1.682  |
| A.4  | 0.056 | 0.084 | 0.107 | 0.316 | 0.595  | 1.000  |

Table 2. Speedup factors between all pairs of CPU implementations using 1 core

The compiler's implicit optimizations provided roughly a 1.5x speedup for the original CPU code (A.1a to A.1b). The code with basic explicit optimizations has a larger 2.9x speedup from compiler optimizations (A.2a to A.2b), partly due to our efforts to help the compiler implicitly vectorize some of the MT19937 random number generation in A.2b, as described in Section 3. While beneficial in the C++ code, this speedup from compiler optimizations is not applicable to the faster vectorized versions implemented in assembly language (A.3 and A.4). The A.1b row of Table 2 is visualized in Figure 15.

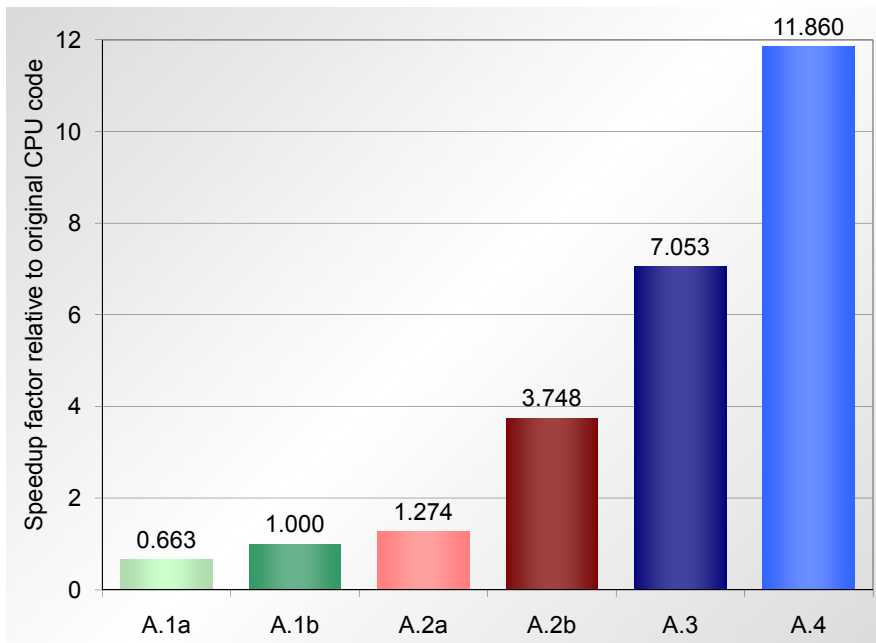

Figure 15. Speedup factors for CPU implementations using 1 core (colours as in Figure 13)

# 5 Conclusion

We presented a simple Metropolis Monte Carlo sweep algorithm and walked through several optimizations on both CPU and GPU, independent of multi-threading. We found that the presented optimizations can gain 9x to 12x speedup for the CPU. Explicitly vectorizing the CPU implementations contributed a 3x speedup and the equivalent changes to allow for memory coalescing on the GPU resulted in a 7x speedup. Moreover, an Intel Core i7 CPU outperforms an NVIDIA GTX-285 GPU by a factor of 2 when comparing our final optimized CPU implementation (A.4) against our final optimized GPU implementation (B.2).

Based on these results, relying solely on compiler optimization for a significant speedup may not be sufficient where high performance is critical, so it is recommended that explicit software optimization, especially vectorization, be more thoroughly examined, tested, and applied. It is also recommended that comparisons of CPU performance against GPU performance utilize the CPU's full vectorization capabilities, if applicable, since neglecting these capabilities could lead to biased comparisons.

## Acknowledgments

We would like to thank Geordie Rose and Mohammad Amin for valuable assistance on this project. We would also like to thank Theirry Drumel for running our GPU performance tests, and the volunteers contributing computer time to AQUA@Home for encouraging us to push the limits of performance.

## Appendix: Derivation of Exponential Approximation

The approximations presented in Section 2.4 is dependent on the IEEE 32-bit floating-point format [8]:

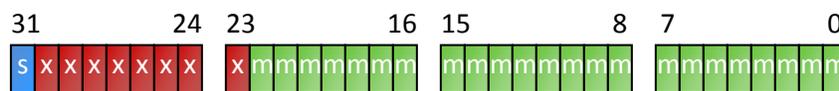

Figure 16. IEEE 32-bit floating-point format (s=sign, x=exponent, m=mantissa)

Values in this format can be denormal ($x=0$), normal ($0<x<255$), infinite ($x=255$ and $m=0$), or NaN ($x=255$ and $m\neq 0$). The value of a normal floating-point number represented by sign $s$, exponent $x$, and mantissa bits $m$, in this format is:

$$f = (-1)^s \left(1 + \frac{m}{2^{23}}\right) 2^{x-127}$$

Supposing that $f$ is positive, i.e. $s=0$, the value of the 32-bit integer that has the same bit representation as this floating-point number is given by:

$$i = 2^{23} x + m$$

Assuming $x<254$, adding $2^{23}$ to $i$ will increase the value of $x$ by one, doubling $f$. This can be repeated until $x=255$, at which point $f$ will become infinite or NaN. Because adding a constant to $i$ has the effect of multiplying $f$ by 2, it is clear that in some form, the value of $f$ is exponential in the value of $i$, and because $f$ and $i$ have the same bit representation, no operation is required for this exponential to be evaluated. Letting:

$$y = \frac{i}{2^{23}} - 127,$$

then isolating the exponent and mantissa of $f$ for a given value of $i$:

$$x = \left\lfloor \frac{i}{2^{23}} \right\rfloor = \lfloor y \rfloor + 127; \quad m = i \bmod 2^{23} = 2^{23}(y \bmod 1); \quad s = 0$$

$$f(i) = (-1)^s \left(1 + \frac{m}{2^{23}}\right) 2^{x-127} = \left(1 + \frac{2^{23}(y \bmod 1)}{2^{23}}\right) 2^{\lfloor y \rfloor + 127 - 127} = (1 + y \bmod 1) 2^{\lfloor y \rfloor}$$

This is exactly $2^y$ when y is an integer, and otherwise, it is a linear interpolation between those values. Thus, given a value of y for which to compute an approximate value of $2^y$, one can instead compute the 32-bit integer:

$$i = \text{round}(2^{23}(y + 127)),$$

and use the resulting 32 bits as a 32-bit floating-point number. Integrating the relative error of this approximation from 0 to 1 yields that the average relative error is $(2\ln^2 2)^{-1} - 1 \approx 0.0407$. Therefore, multiplying the result by $2\ln^2 2$ gives a relative error averaging to zero. The relative error is plotted in Figure 17. Note that this can produce unpredictable results when y<-126 or y≥128.

Further improvement of this approximation is based on that while the approximation of $2^y$ is exact when y is a multiple of 1, the approximation of $2^{4y}$ is exact when y is a multiple of 0.25, so the latter approximation is exact with 4 times the frequency. Given a method of quickly approximating 4th roots with lower error than this approximation, such as with special SSE instructions for approximate reciprocal-square-roots, one can compute:

$$i_4 = \text{round}(2^{23}(4y + 127))$$
$$2^y \approx \sqrt[4]{2\ln^2 2 \, f(i_4)}$$

This has a relative error roughly bounded by (-0.01, 0.005), more accurate but slower than the previous method. This relative error is also shown in Figure 17. Note that to avoid overflows, one must have -31.5≤y<32 for this more accurate approximation.

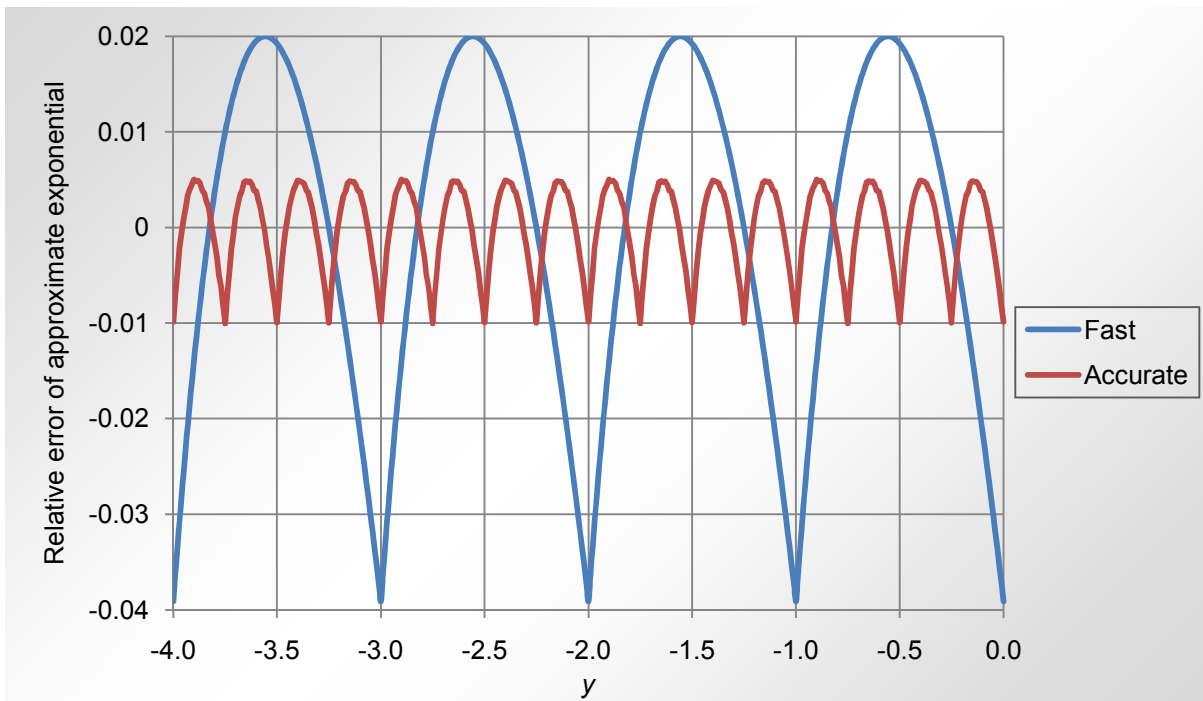

Figure 17. Relative error of fast and accurate exponential approximations by input value